\documentclass[10pt,conference]{IEEEtran}
\IEEEoverridecommandlockouts
\usepackage[]{collab}
\usepackage{enumitem}
\usepackage{balance}
\usepackage{xcolor}
\usepackage{setspace}
\usepackage{subfigure}
\usepackage{multirow}
\usepackage{multicol}
\usepackage{tabularx}
\usepackage{siunitx}
\usepackage{float}
\usepackage{color}
\usepackage{hyperref}
\usepackage{diagbox}
\usepackage{caption}
\usepackage{makecell}
\usepackage{colortbl}
\usepackage{tcolorbox}
\usepackage{mdframed}
\usepackage{listings}
\usepackage{booktabs}
\usepackage[utf8]{inputenc}
\usepackage{calligra}
\usepackage[normalem]{ulem}
\usepackage{indentfirst}
\usepackage{url}
\usepackage{soul}
\usepackage{nicematrix}
\usepackage{fontawesome5}
\usepackage{textcomp}
\usepackage{stfloats}
\usepackage{verbatim}
\usepackage{graphicx}
\usepackage{amsmath,amsfonts}
\usepackage[linesnumbered,ruled,vlined,noend]{algorithm2e}
\usepackage{array}
\usepackage[utf8]{inputenc}

\collabAuthor{jj}{blue}{Junjie}
\collabAuthor{gb}{magenta}{Guangba}
\collabAuthor{zh}{purple}{Zhihan Jiang}
\collabAuthor{jz}{magenta}{Jiazhen Gu}
\collabAuthor{zb}{teal}{Zhuangbin}
\collabAuthor{jc}{googlered}{JC}
\collabAuthor{yj}{orange}{Yujie Huang}
\collabAuthor{yl}{orange}{Yulun Wu}

\newcommand{\boxmargin}{1mm}

\newtcolorbox{myboxa}[2][]{
    colback=gray!10!white,
    colframe=black, enhanced,
    attach boxed title to top left={yshift=-2mm,xshift=5mm},
    title=#2,#1
}
\newtcolorbox{myboxb}[2][]{
    boxsep=3pt,
    left = \boxmargin, right = \boxmargin, top = \boxmargin, bottom = \boxmargin,
    title={#2},#1
}
\newtcolorbox{myboxc}{
    colback=gray!15!white,
    arc = 0pt, outer arc = 0pt,
    boxsep=0pt, left = 3pt, right = 0pt, top = 0pt, bottom = 0pt, 
    leftrule=3pt, bottomrule=0pt,toprule=0pt, rightrule=0pt,
    left = \boxmargin, right = \boxmargin, top = \boxmargin, bottom = \boxmargin
}
\newtcolorbox{myboxd}{
    colback=gray!10,
    colframe=black,
    width=\columnwidth,
    arc=1mm, auto outer arc,
    boxrule=0.5pt,
}

\definecolor{myyellow}{HTML}{FFF2CC}
\newcounter{finding}

\definecolor{myyellow}{HTML}{FFF2CC}
\newcounter{insight}

\newcounter{challenge}

\definecolor{mygreen}{HTML}{AFCFA5}
\newcounter{opportunity}

\begin{document}

\thispagestyle{plain}
\pagestyle{plain}

\title{The Multi-Agent Fault Localization System Based on Monte Carlo Tree Search Approach}


\author{Rui Ren \\ Institute of Computing Technology, Chinese Academy of Sciences; University of Chinese Academy of Sciences \\ Beijing, China \\ renruirui1234@gmail.com}

\maketitle

\begin{abstract}
In real-world scenarios, due to the highly decoupled and flexible nature of microservices, it poses greater challenges to system reliability. The more frequent occurrence of incidents has created a demand for Root Cause Analysis(RCA) methods that enable rapid identification and recovery of incidents. Large language model (LLM) provides a new path for quickly locating and recovering from incidents by leveraging their powerful generalization ability combined with expert experience. Current LLM for RCA frameworks are based on ideas like ReAct and Chain-of-Thought, but the hallucination of LLM and the propagation nature of anomalies often lead to incorrect localization results. Moreover, the massive amount of anomalous information generated in large, complex systems presents a huge challenge for the context window length of LLMs. To address these challenges, we propose \textbf{KnowledgeMind}, an innovative LLM multi-agent system based on Monte Carlo Tree Search and a knowledge base reward mechanism for standardized service-by-service reasoning. Compared to State-Of-The-Art(SOTA) LLM for RCA methods, our service-by-service exploration approach significantly reduces the burden on the maximum context window length, requiring only one-tenth of its size. Additionally, by incorporating a rule-based real-time reward mechanism, our method effectively mitigates hallucinations during the inference process. Compared to the SOTA LLM for RCA framework, our method achieves a 49.29\% to 128.35\% improvement in root cause localization accuracy.
\end{abstract}


\section{Introduction}

Monolithic services have gradually been transformed into more granular modules, consisting of hundreds or even thousands of loosely coupled microservices~\cite{seer, deathstar, suite, intro1}. Prominent companies such as Netflix, eBay, and Alibaba have embraced this application model. Microservices bring a range of benefits, including simplifying application development, improving resource provisioning efficiency, and offering greater flexibility. However, despite these advantages, microservices can introduce complex interactions between modular services, making on-demand resource provisioning more difficult and potentially causing performance degradation. 

Not only that, but the microservice architecture makes the business logic of individual services more flexible and convenient. Business logic can be implemented and executed in various languages such as Java, Python, Go, etc. While this enhances developer productivity, it also poses a greater challenge for operations. SREs(Site reliability engineering) need to be familiar with the code logic implemented in different languages and identify specific root causes. According to Google's white paper~\cite{ver3}, over 70\% of failures occur after changes to service code. 

To help engineers resolve failures efficiently, fault localization plays a central role in maintaining online service systems~\cite{icse3,icse4,icse5,icse9}. With the development of monitoring and collection tools, Metrics, Traces, and Logs have emerged as the three essential components of fault localization. Metrics are numerical data collected at regular intervals, offering insights into the underlying causes of an application’s behavior. Each trace captures the sequence of a request as it moves through service instances and their operations~\cite{intro7,intro8}. Logs deliver in-depth information about the system’s status and user interactions.

With the rapid iteration and development of large language models (LLM), many researchers have begun building various operational capabilities for microservice systems based on LLM. Compared to other methods, LLMs offer significant advantages in areas such as code parsing and log analysis, providing better generalization than traditional rule-based and small-model algorithms(e.g. LSTM, GCN and CNN)~\cite{llm3,llm4,llm5}. LLMs do not require extensive expert knowledge to define specific rules for service monitoring information. When an abnormal log appears, they can autonomously and accurately analyze and summarize key information by combining the relevant code, without the need for supervision. Additionally, LLM can serve as a central strategy hub, calling tools and analyzing the next steps in the strategy. They enable unified scheduling and joint analysis of multiple data sources, while also flexibly exploring the entire system in an effective manner~\cite{llm6,llm7}.

\section{BackGround and Motivation}\label{sec:related_work}

\subsection{Basic Concepts of Microservice System}

\begin{figure}
    \centering
    \includegraphics[width=1.05\linewidth]{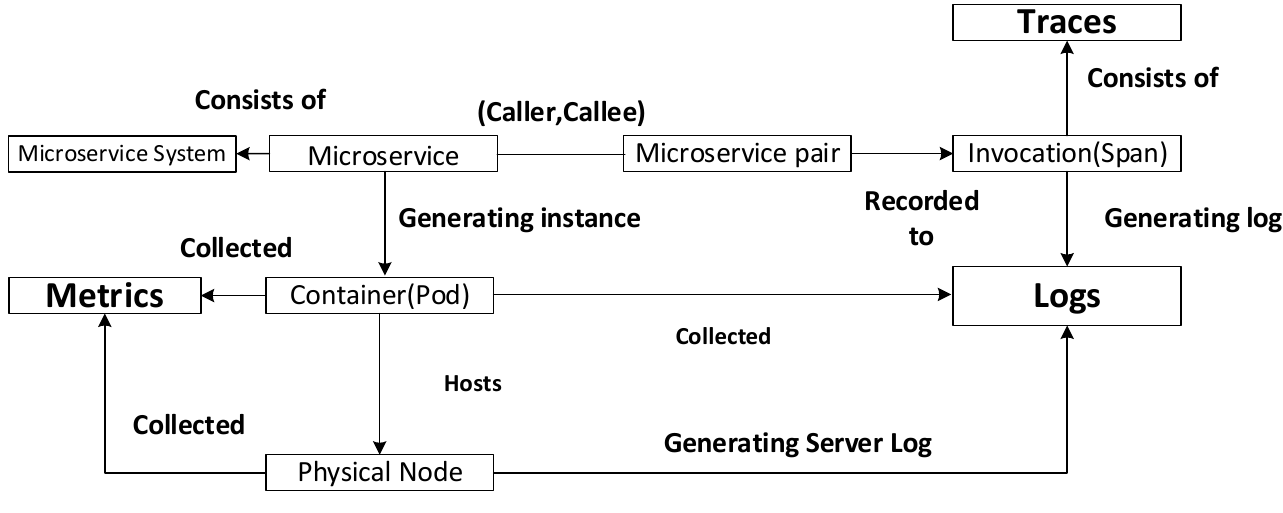}
    \caption{The Overview of Microservice System.}
    \label{fig:basic}
\end{figure}

In Figure~\ref{fig:basic}, we show the basic concepts of the microservice system. A \textbf{Microservice System} is a highly efficient architecture comprising of \textbf{Microservices} that are loosely coupled and lightweight. \textbf{Microservices} are initiated by \textbf{instances}, which is the running version of this microservice on the \textbf{Physical Node}. The status of both the \textbf{Physical Nodes} and the \textbf{Instances} are described through \textbf{metrics} generated by monitoring tools such as Prometheus~\cite{Prometheus}. The \textbf{trace} describes the execution process of a request through service instances~\cite{elasticsearch}. It has been supported by open-source solutions~\cite{jaeger,opentelemetry}. The \textbf{Logs} are produced during each service invocation, which is generated by the logging statements and records the internal status of the invocation.

\subsection{Prior Work of LLM for AIOps}

Some existing research have focused on utilizing LLMs to accomplish fault localization and anomaly detection, such as monitorAssistant~\cite{monitorassistant}, RCACopilot~\cite{rcacopilot}, RCAgent~\cite{rcagent} and mABC~\cite{mabc}. monitorAssistant aims to use LLMs for tasks like alert ranking and generating detection reports, while RCACopilot leverages LLMs' summarization capabilities to construct case libraries for new fault localization. RCAgent, on the other hand, focuses on recursively mining and summarizing effective information from logs using LLMs to achieve fault localization. mABC designed a multi-agent LLM RCA framework, which uses a multi-agent collaboration model to perform RCA. It leverage the ReAct~\cite{react} framework to breaks down tasks into sub-tasks, such as decoupling the analysis of abnormal metrics, fault topology, and other information through multiple agents. It also uses a judge agent to determine whether there are anomalies in the results analyzed by the multiple agents. This approach can effectively enhance the accuracy of LLM in RCA scenarios. However, mABC only decomposes the preceding tasks, and the final root cause determination still involves aggregating all the analyzed information from the system and using a single-step judge agent for judgment. In cases of complex fault topologies, faults may propagate, making it difficult for the judge agent to identify the true root cause from multiple anomalous services. 
Moreover, the ReAct process cannot guarantee the correctness and accuracy of the exploration, especially in the absence of expert knowledge guidance. As a result, mABC can easily produce irrelevant information during the exploration process, leading to hallucinations in the large model. Beside, in large systems, the ReAct's data generated by multiple agents may exceed the maximum token length.

Although previous works has proposed new approaches for using LLM in operations and provided various effective solutions, these approaches still face many challenges.

\textbf{Challenge 1: Unconstrained Stochastic Analysis and Hallucinations in LLM}.
Existing LLMs are autoregressive probabilistic prediction models. As a result, different pre-trained models analyze anomalies during operations based on their own pre-trained knowledge combined with the current situation. This often leads to hallucinations and random analytical results that are unrelated to the actual operations scenario, making it difficult to provide operations personnel with effective reasoning and analysis.

\textbf{Challenge 2: Fault Propagation Could Lead to Incorrect Judgments and reasoning result}.
Fault propagation can lead to abnormal behaviors across multiple services. The LLM will perform information summarization and root cause analysis for multiple abnormal services, but without effective guidance from the current system's rules or case libraries, it is prone to hallucinations. The model may generate multiple analysis results based on these anomalies, all of which might appear valid and reasonable. However, this can greatly hinder operations personnel in accurately determining the root cause.

\subsection{Motivations} 
The existing works of LLM for RCA~\cite{monitorassistant,rcagent,rcacopilot,llm1,llm2} mostly focus on leveraging the powerful natural language analysis of LLMs to summarize alarms and fault results, or simply feeding all data into the LLM in natural language form for black-box analysis. This keeps the process opaque and hard to understand. Additionally, directly summarizing and outputting all analysis results makes it difficult to determine whether there are errors in the reasoning process of the LLM. This leads to the issue where the LLM may provide different fault localization results for the same problem in multiple question-answering sessions. Therefore, we aim to white-box and standardize the reasoning process. It could ensure the stability and reliability of the reasoning results as much as possible.

Therefore, we propose a knowledge-driven, Monte Carlo Tree Search based LLM multi-agent root cause analysis system. This system standardizes each step of the reasoning process by constructing a fault reasoning tree, aiming to reduce the randomness and hallucinations that may exist in current LLM Agent for aiops framework. Additionally, it leverages rule-based rewards from the knowledge base to build a Monte Carlo Tree Search (MCTS) process, enhancing the effectiveness and accuracy of reasoning. By leveraging agents for logs, metrics, and traces, the system fully utilizes valuable information from each service. Through effective orchestration of this information, combined with rules and cases from the knowledge base, the system achieves accurate and efficient identification of root cause services and types.

Our contributions could be summarized as follows:

\begin{itemize}
    \item[1)]  We propose the KnowledgeMind framework, the first multi-agent fault localization system based on a Monte Carlo Tree Search(MCTS) process. In this framework, compared to other baselines that leverages original ReAct and Chain-of-Thought (CoT) to finish RCA, we avoid the inference procedure's hallucination and ensure the effectiveness of each reasoning step through knowledge base rule-based rewards and a service-by-service exploration process. Moreover, this method reduces the reliance on the maximum token limit for LLM and minimizes the risk of incorrect fault localization caused by fault propagation.
    
    \item[2)] We introduce the Fault Mining Tree for the first time, combining it with Monte Carlo Tree Search (MCTS) to implement a service-by-service reasoning process. The Fault Mining Tree helps constrain the potentially excessive search space in microservice system, enabling the LLM to complete the search and analysis of new services in each round. This also helps large models better understand the impact of graph structures on anomaly propagation's system and accurately pinpoint the root cause in the relevant service.
   
    \item[3)] We have innovatively proposed a multi-agent collaboration system that combines both LLM and traditional models. The Metric and Log Agents enhance the system's diagnostic capabilities by integrating the traditional algorithms. Additionally, we introduce the Knowledge Base Agent and Verifier Agent to collaborate in improving the accuracy of each step in the reasoning process.

    \item [4)] On real-world system public datasets, compared to the current state-of-the-art LLM Agent for RCA framework, KnowledgeMind improves the accuracy of fault service and fault type localization by 49.29\% -- 128.35\%. 
\end{itemize}


\section{The Knowledge Mind}

\subsection{The Pipeline of KnowledgeMind}
\begin{figure*}[h]
    \centering
    \includegraphics[width=1\linewidth]{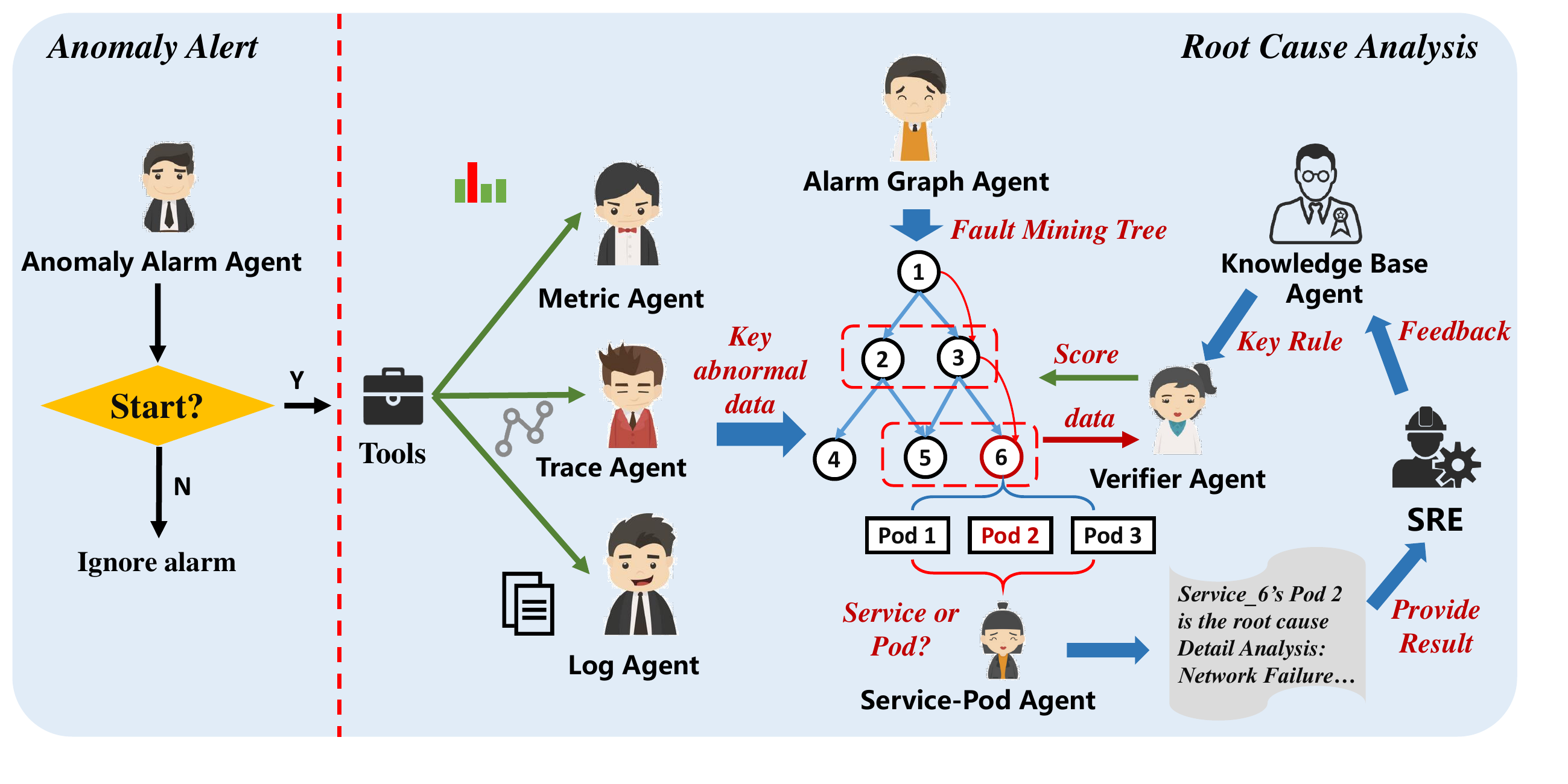}
    \caption{The Pipeline of KnowledgeMind.}
    \label{fig:klm}
\end{figure*}

the KnowledgeMind is an end-to-end fault localization framework for microservices system based on multi-agent and LLM. As shown in Figure~\ref{fig:klm}, the Anomaly Alarm Agent first detects the existing anomalies in the current system and provides the anomaly information to the LLM to decide whether to enable the RCA process. If the decision is made to enable it, the Alarm Graph Agent constructs the alarm propagation topology and generates the Fault Mining Tree. Based on the Fault Mining Tree, Monte Carlo Tree Search (MCTS) is used to perform step-by-step reasoning. During each reasoning step, the Metric, Log, and Trace Agents are called to gather the necessary anomaly information. The Verifier Agent, combined with the Knowledge Base Agent, uses case library and expert rules to determine the score of the current reasoning process and updates the Monte Carlo Tree. After several rounds of iteration, the final root cause is identified by the highest score, indicating the service where the root cause is located. Once the root cause service is determined, the Service-Pod Agent is used to further identify whether the root cause is related to the entire service or a single pod of the service, thus refining the granularity of the fault location.

\subsubsection{Log Agent}
Due to the token limitations of LLMs and the fact that logs consume significantly more tokens compared to metrics and other data, as illustrated in the figure, we first introduce a Log Agent to perform preliminary filtering and extraction of information before proceeding with step-by-step reasoning.

Our Log Agent operates in two main steps. First, it performs an initial extraction and analysis of key information based on Drain and keywords. Then, it conducts more refined filtering and extraction using clustering methods and knowledge base matching.

Drain~\cite{drain} is a template extraction method based on a template tree, designed to quickly record and process all encountered log templates. Using the templates extracted by Drain, along with the corresponding log entries recorded in those templates, we set specific keywords to filter which templates should be retained and prioritized for analysis. For example, in the AIOPS 2022 Dataset~\cite{AIOPS2022}, templates containing keywords such as "exception" and "error" are identified as high-priority targets. The associated log entries are preserved for further processing.

Furthermore, if the number of processed log entries falls within the token limits of the LLM, we directly pass them to the Log Agent for analysis and summarization. If the entries still exceed the token limits, clustering or knowledge-base matching is employed for a second round of filtering. We provide two different filtering methods to address the system's cold-start problem. When the system lacks a knowledge base(cold-start), clustering can effectively deduplicate similar logs. However, when a knowledge base is available, we aim to leverage more accurate expert knowledge to complete the deduplication process. The clustering leverages the Gaussian Mixture Model(GMM)~\cite{gmm} combined with TF-IDF~\cite{tfidf} to further cluster key log information, and then select all the information from the anomalous log categories. The knowledge-base matching retrieves expert-summarized anomalous template information to perform secondary selection on the current templates, or uses expert-summarized normal template information to exclude the current templates.

After completing the screening of key log entries, we allow the LLM to further explore the code context that prints the current log by combining code queries with recursive indexing. It recursively analyzes the related functions within the context until the model determines that all valuable information has been searched and analyzed. Finally, we send all the critical information to the LLM for a summarization, which serves as the ultimate output of the Log Agent.

\subsubsection{Metric Agent}
Due to the inherent limitations of LLMs in effectively understanding time-series data, we designed a Metric Agent that incorporates various time-series anomaly detection algorithms to help the LLM comprehend the key information embedded in the metrics. Taking single-metric detection as an example, we use the ARIMA algorithm to analyze whether each metric exhibits abnormal states. First, the model's predicted metric values $\hat{y}_t$ are obtained based on ARIMA. Further, we calculate the residuals by comparing the actual values with the predicted values based on formula~\ref{eqn:residual}. Where $y_t$ represents groundtruth, $e_t$ represents the calculated residual, $z_t$ denotes the standardized residual, and $\lambda$ is the anomaly detection threshold set based on expert experience. Using expert experience, threshold values for detecting abnormal spikes and dips are set, and the residuals are compared with these thresholds. If the residuals exceed the threshold, a corresponding statement, such as 'Timestamp: SPIKE of metric\_name' or 'Timestamp: DIP of metric\_name', is output in natural language. The Metric Agent can also analyze multivariate correlations using deep learning-based anomaly detection methods(e.g., LSTM, GRU), presenting the results as "anomaly existing" or "anomaly unexisting" to the LLM.

\begin{align}
\begin{split}
\label{eqn:residual}
    e_t &= y_t - \hat{y}_t \\
    z_t &= \frac{e_t}{\hat{\sigma}_e} \\
    |z_t| > \lambda\\
\end{split}
\end{align}

\subsubsection{Trace Agent}
The Trace Agent does not involve graph-related analysis. The Trace Agent primarily focuses on analyzing the invocation latency and communication status of individual services. It is used to detect whether there is an abnormal increase in invocation latency or service invocation failures within the current time window. Invocation latency detection related to Trace still relies on the anomaly detection methods mentioned in the Metric Agent to determine abnormal states.

\subsubsection{Anomaly Alarm Agent}

The Anomaly Alarm Agent relies on the detection capabilities mentioned in Trace, Log, and Metric Agent to issue alarms for abnormal system states. This agent is solely used to determine which services in the system have alarms and whether the fault localization process needs to be started. Compared to the detailed analysis processes of the Trace, Log, and Metric Agents, it focuses more on the rapid analysis and processing of real-time data.

For example, to detect alarms in logs, we use a keyword-matching approach to determine whether any services in the system have alerts (e.g., continuous occurrences of 'error' or 'exception'). However, it is worth noting that, unlike the Log Agent, which performs detailed filtering and analysis on the entire set of logs, we directly determine whether fault localization is needed upon detecting abnormal information.

\subsubsection{Knowledge Base Agent}
The Knowledge Base Agent is used to store expert rules and the case library. The expert rules guide the scoring process of the verifier agent, helping the LLM assess the anomaly level of a set of services in an unsupervised state. 

The case library is used to store complete anomaly information and analysis results of each service after an SRE confirms that KnowledgeMind has accurately located a fault case. In the procedure of inference, the case library identifies the top-k cases most similar to the current service anomalies based on Jaccard similarity. Subsequently, it performs a secondary matching by comparing the abnormal information of already explored services with the top-k cases to determine the fault case most similar to the current analysis scenario. Finally, leveraging the LLM to evaluate whether the most similar case in the knowledge base fully matches the current fault to determine the root cause service, fault type, and solution. 

These cases can speed up the MCTS search process because when the LLM determines that a case completely matches the current search result, the search process ends immediately, returning the analysis results from the case library.

\subsubsection{Alarm Graph Agent}
The Alarm Graph Agent first collects all service call relationships to construct a complete service dependency graph. Then, it uses the alarm service set provided by the Anomaly Alarm Agent to depict the specific alarm propagation graph. The process is shown in the pseudocode~\ref{alg:alarm_topo} below.

\begin{algorithm}[!htbp]
\caption{Alarm Topology Extraction}\label{alg:alarm_topo}
\KwIn{Service dependency graph $G_{\text{dep}}$, set of services with alarms $N = \{M_1, M_2, ..., M_k\}$}
\KwOut{Real alarm propagation topology $G$}

Extract all alarm services and their ancestor services from $G_{\text{dep}}$ to generate a potential propagation topology $G'$ \;
$G' \gets$ ExtractAncestorServices($G_{\text{dep}}$) \;

Compute the intersection of services in $G'$ and the services with alarms in $N$ \;
$G \gets G' \cap N$ \;

\Return $G$ \;
\end{algorithm}

\begin{figure}[h]
    \centering
    \includegraphics[width=1\linewidth]{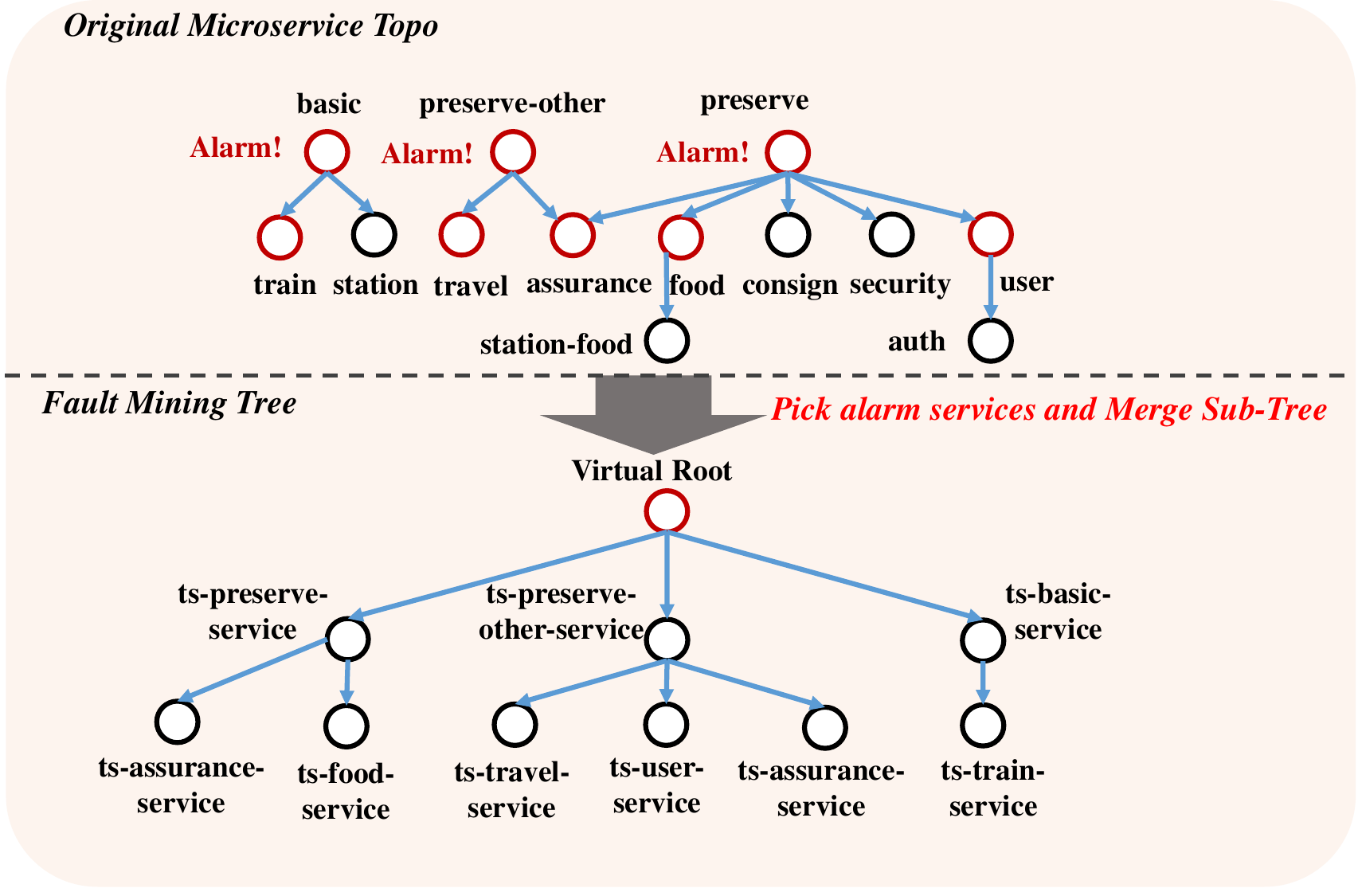}
    \caption{The Construction of Fault Mining Tree.}
    \label{fig:fmt}
\end{figure}

As shown in the figure~\ref{fig:fmt}, the Alarm Graph Agent completes the construction of \textbf{Fault Mining Tree}. Failures in underlying microservices may cause anomalies to propagate across multiple regions. The microservices in these regions might not necessarily be invoked by the same services, thereby forming multiple anomalous subgraphs. Therefore, we decompose this original alarm graph into multiple tree-structured subgraphs based on services' dependent relationship, named \textbf{Sub-Tree}. Please note that the Sub-Tree is constructed based on the service granularity, meaning that we do not distinguish between individual pods. Any anomalies present in all pods of a service are considered as anomalies of the same service. 


Next, we connect the root nodes of these tree structures to the same root node (referred to as the \textbf{virtual\_root}). This method allows all potential invocation paths to begin fault analysis from a single entry point, making it easier to obey the Monte Carlo Tree Search process. We named the newly generated dependency structure with the virtual\_root node as the \textbf{Fault Mining Tree}.

\subsubsection{Verifier Agent}
As the figure~\ref{fig:klm} shown, the verifier agent simultaneously compares the anomaly information of all services under the same root in the Fault Mining Tree to generate a ranked score. These scores are used in the subsequent iterative exploration process to select the highest-scoring nodes for completing the simulation. Concurrently, the Verifier Agent incorporates a set of pivotal rules based on expert experience to facilitate the scoring process. For example, if severe error alarms are present in the logs of the current service, a higher score is assigned, while warning alarms will receive a lower score. The Verifier Agent analyzes the informations from its internal rules with the feedback provided by the Metric, Log, Trace. This consolidated information is then relayed to the LLM, which is tasked with delivering the ultimate assessment score (ranging from 0 to 10).

\subsubsection{Fault Reasoning Step-By-Step}

\begin{figure*}[h]
    \centering
    \includegraphics[width=1\linewidth]{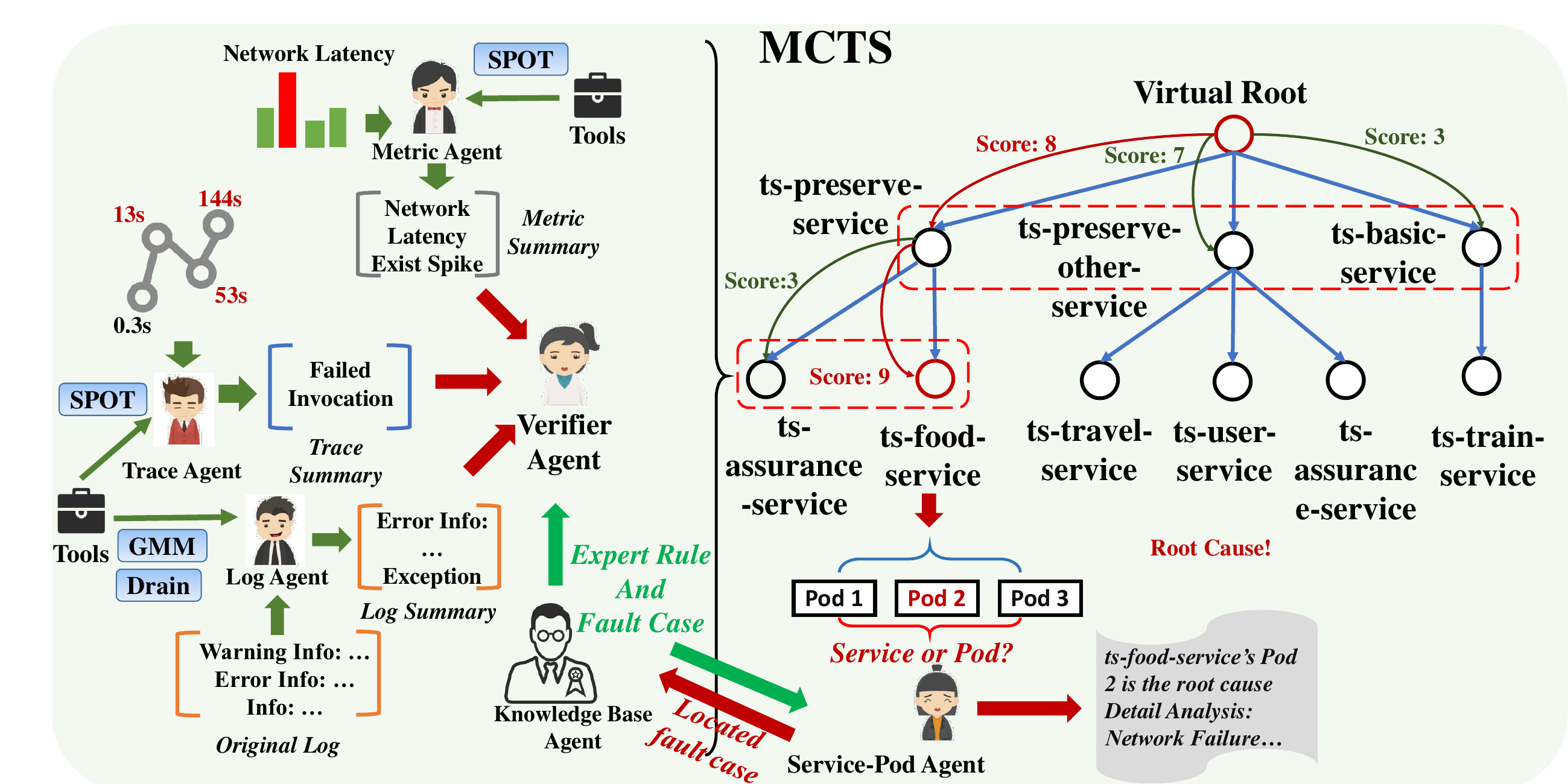}
    \caption{The Procedure of Fault Reasoning Step-By-Step.}
    \label{fig:MCTS}
\end{figure*}

Unlike previous studies, we abandoned the traditional approaches that analyze all anomaly and alarm data directly. As shown in Figure~\ref{fig:MCTS}, we perform a service-by-service search on the Fault Mining Tree using MCTS. Each search step relies on key anomaly information provided by the Metric, Trace, and Log Agents, and the Verifier Agent is used to evaluate and select the service with the highest abnormal score at the current layer. Next, we will provide a more detailed explanation of the MCTS reasoning process on the Fault Mining Tree through pseudocodee~\ref{alg:MCTS}.

The inference begins from the virtual\_root, evaluating the first layer of each subtree that involves services. Initially, the LLM employs Log, Trace and Metric Agents to retrieve key alarm information for every SubTrees' service. The alarm information is then evaluated by the Verifier agent, which calculates the anomaly scores of all direct descendants under the current parent node. Note that the Verifier agent is implemented using an LLM combined with specific scoring rules. Using the Verifier agent, we could get the score of every child node(service). Once the Best child is identified, we further perform simulation using the greedy policy based on the Knowledge Base.

It is important to note that this simulation is not similar to the traditional Monte Carlo Tree Search (MCTS) where exploration continues step by step down the tree until the terminal. Due to the potential complexity of service dependencies and the fixed exploration path of the Fault Mining Tree, performing deeper exploration into each subtree at every simulation step to summarize and analyze the alarms from all lower-level services would not only make the LLM prone to hallucinations but also poses challenges to the maximum input token limit of the LLM. 

Therefore, we designed a new strategy to complete the simulation procedure. First, we use the Knowledge Base Agent to search for the case most similar to the current service anomaly from the case library and pass it to the LLM for matching. If the LLM determines that the current case perfectly matches the anomaly, the search is immediately terminated, and the root cause details and fault resolution from the Knowledge Base are returned. If the LLM finds no successful match, it then explores the abnormal status of all services dependent on the current service(direct ancestor services) and assigns scores based on their abnormal status. This is done because when a root cause service fails, all services that call it will also enter an abnormal state due to its failure.

Completing the above process, we enter the $BackPropagation()$ function. During this phase, we update the evaluation values of all related nodes and proceed to the next iteration. After multiple rounds of MCTS exploration, we identify the most frequently visited fault path along with the corresponding root cause service and fault type analysis.


\begin{algorithm}[!htbp]
\caption{Monte Carlo Tree Search For Fault Mining Tree}
\label{alg:MCTS}
\KwIn{$s_0$: Initial state, $N$: Number of simulations}
\KwOut{The best action $a^*$ from the root node}
\SetKwFunction{UCT}{UCT}
\SetKwFunction{Expand}{Expand}
\SetKwFunction{Simulate}{Simulate}
\SetKwFunction{Backpropagate}{Backpropagate}
\SetKwFunction{BestChild}{BestChild}
\SetKwFunction{GetValidActions}{GetValidActions}
\SetKwFunction{TakeAction}{TakeAction}
\SetKwFunction{Fn}{Function}
\textbf{Initialization:} Create the Virtual\_Node $v_0$ for $s_0$ \;
\For{$i \gets 1$ to $N$}{
    \textbf{Selection:} Start at $v_0$ and recursively select child nodes using \UCT(v) until the unexpanded node $v_l$ is reached\;
    \textbf{Expansion:} If $v_l$ is not a terminal node and its child nodes $C=\{v_{c1},...,v_{ck}\}$ have not been explored, these child nodes corresponding to services are explored through \Expand($v$)\; 
    \textbf{Simulation:} From $v_c$, simulate a greedy rollout to obtain a reward $R$ using \Simulate($v$)\;
    \textbf{Backpropagation:} Backpropagate the reward $R$ from $v_c$ to $v_0$, updating visit counts and value estimates using \Backpropagate($v, R$)\;
}
\textbf{Output:} Return the best action $a^*$ from the root node $v_0$ using the highest visit count\;
\Fn{\UCT($v$)}{
    \Return $\frac{Q(v)}{N(v)} + C \sqrt{\frac{\log N(parent(v))}{N(v)}}$ \;
}
\Fn{\Expand($v$)}{
    \textbf{Step 1:} Abnormal information mining is performed on all child nodes $C=\{v_{c1},...,v_{ck}\}$ using the Metric, Trace, and Log Agents.\;
    \textbf{Step 2:} Evaluate all unexplored child service $C=\{v_{c1},...,v_{ck}\}$ of current service and generate their abnormal score $S=\{s_{c1},...,s_{ck}\}$ by Verifier Agent\;
    \textbf{Step 3:} Calculate the highest score of all children service $s_h$ in $S$\;
    \Return $(s_h,v_h)$;
}
\Fn{\Simulate($s_h,v_h$)}{
    \textbf{Step 1:} Check if knowledge base can determine the root cause from $(s_h,v_h)$\;
    \If{KnowledgeBase can determine the root cause}{
        \Return the reward $R$ based on root cause detection\;
    }
    \Else{
        \textbf{Step 2:} Analyze the anomaly count of all services that call the current service $v_h$\;
        Let $C = \{v_{c1}, v_{c2}, ..., v_{ck}\}$ be the services that call $v_h$\;
        Calculate the anomaly count for each $v_{ci}$, denoted as $A(v_{ci})$\;
        Calculate the final state score $S$ based on the total anomaly count of all services in $C$\;
        \Return the reward $R$ based on score $S$\;
    }
}
\Fn{\Backpropagate($v, R$)}{
    \While{$v \neq \text{null}$}{
        Increment $N(v)$\;
        Update $Q(v) \gets Q(v) + R$\;
        $v \gets parent(v)$\;
    }
}
\end{algorithm}



\subsubsection{Service-Pod Agent}

The above search process completes the fault reasoning at the service level. We then further analyze the anomaly situations across all pods of the current service. If all anomalies are concentrated in a single pod, the root cause is determined to be a pod-level fault. If multiple pods fail simultaneously, the root cause is identified as a service-level fault.

SRE can explicitly understand the entire reasoning process of KnowledgeMind to determine whether the root cause path, root cause service, and fault reason have been accurately identified. If any errors arise in the reasoning process due to issues with rules or the case library, SRE can correct the expert rules and reanalyze the fault. Further elaboration on these details of prompt template of Multi-Agent can be found in the appendix~\ref{app:A}.

\section{Experiments}\label{sec:exp}

In this section, we perform experiments on the open-source datasets to show the performance of KnowledgeMind in comparison with the state-of-the-art fault localization algorithms and tested the performance of KnowledgeMind when integrated with different LLMs.


\subsection{Experimental Setup}

\subsubsection{Datasets} We conduct experiments on two public datasets, which are denoted as dataset $\mathcal{A}$ and dataset $\mathcal{B}$.

Dataset $\mathcal{A}$ comes from the AIOPS 2022 datasets. This dataset is sourced from a real system of one of the largest banks in China. The system includes 43 services, with a total of 167 faults occurring over four days. Dataset A contains over 60GB of Trace, Metric, and Log data. The collected metrics include more than 250 types, the trace records detailed information about the latency of all request calls, and the logs provide a comprehensive display of all the printed information during services' running. The fault types are shown in Figure~\ref{tab:fault_type}, covering aspects such as CPU, memory, network, Disk I/O exhaustion, and process termination. The fault hierarchy includes individual service pods, all pods of a service. 

Dataset $\mathcal{B}$ are obtained from the Trainticket benchmark platform~\cite{trainticket}. The system consists of 41 microservices, with 165 faults injected across different services, including CPU exhaustion, network delay, and container pause faults. The dataset includes trace and metric data, with the metrics covering key information such as memory, CPU, network delay, HTTP status, and file disk I/O.

\begin{table}[htbp]
\caption{Overview fault type of three datasets}
\begin{center}
\begin{tabular}{c c c c}
\hline
\textbf{Dataset}&\textbf{fault type}&\textbf{cases}&\textbf{cases percent}\\
\hline
&File system I/O(2 types) &28&16.8\%\\
&Network problem(4 types)&84&50.3\%\\
A&CPU problem&23&13.8\%\\
&Memory problem&22&13.2\%\\
&Process Pause&10&6\%\\
\hline
&Network-delay&56&34\%\\
B&CPU&51&31\%\\
&Container Pause&58&35\%\\
\hline
\end{tabular}
\vspace{-20pt}
\label{tab:fault_type}
\end{center}
\end{table}

\subsubsection{Evaluation Metrics}
In the root cause localization scenario, SREs typically focus on the specific fault location and fault type~\cite{dejavu,sage,slim,tracerca}. Therefore, we evaluate the accuracy of our algorithm and other baseline based on two metrics. Since KnowledgeMind supports both unsupervised (without case library) and semi-supervised (with case library) modes, we adopt the commonly used fault service location($FL$) and fault type location($FT@k$) in most unsupervised and semi-supervised root cause localization. algorithms~\cite{tracerca,microrank,slim,dejavu} to measure fault location and fault type. Since fault type localization is more challenging than service localization, we ask the model to provide the top 3 root cause results for evaluation. Please note that $FT@k$ calculates the fault type localization accuracy based on accurate service localization.

\subsubsection{Baseline Methods}
The baseline algorithms we use for comparison include COT(Chain-Of-Thought)~\cite{cot}, mABC~\cite{mabc}, and RCAgent~\cite{rcagent}. COT guides the LLM to reason about the specific failure step by step through a corresponding chain-of-thought template. mABC employs a combination of mABC~\cite{react} and an Agent-based approach to enable the LLM to analyze the failure, while Direct directly passes all anomaly information to the LLM for analysis and understanding, producing the final result directly.

Since all these root cause localization algorithms are drived based on LLM, we further test these algorithms on different base models. Besides, to fully analyze the impact of the base LLM on root cause localization performance, we test multiple mainstream MoE(Mixture of Experts) models~\cite{moe} and Dense models on datasets A and B for root cause localization effectiveness. These models include Qwen-QwQ~\cite{qwen2}, LLAMA3.3-70B~\cite{llama}, GPT-4-turbo~\cite{gpt4}, Qwen-Max~\cite{qwenmax} and DeepSeek-V3~\cite{deepseek}.

Because RCAgent is not open-sourced and mABC only aggregates some metric information without providing specific detection methods, this is inadequate for effectively detecting issues in our dataset. To ensure fairness, we have incorporated anomaly detection algorithms similar to those in the Metric Agent and Log Agent into their approach to replicate the experiment accurately. Besides, these methods like RCAgent and mABC do not have a knowledge base for supervised RCA. Therefore, KnowledgeMind only uses rules during the inference process without enabling the case library. This means that there are no historical fault cases in the knowledge base for reference.

\subsection{Performance on Fault Localization}

As shown in Table~\ref{tab:acc}, compared to methods based on Chain-of-Thought templates, mABC, and Direct approaches, our method demonstrates superior analytical performance. This is because Chain-of-Thought and mABC struggle to ensure that each step effectively completes the service detection and analysis. Invalid analytical steps may negatively impact LLM analysis and cause hallucinations. Moreover, LLMs struggle to effectively understand the propagation relationships of anomalies from graph structures, which can lead to incorrect root cause identification.

On dataset A, our method outperforms other baseline models by a significant margin in terms of accuracy. The performance differences of KnowledgeMind across multiple base LLMs are not significant. This is because our evaluation process relies more on multi-agent collaboration combined with expert rules, without overly depending on the reasoning capabilities of the base LLM. For dataset B, our method does not show a significant advantage. This is because the faults injected into the system are relatively simple, the number of metric types is fewer, and the representation of each fault can be easily observed from a single metric. For simple scenarios, methods based on COT and ReAct, such as RCAgent and mABC, can effectively identify the root cause with fewer analysis steps. In complex scenarios with more diverse metric and log data, KnowledgeMind is better able to leverage its advantages.


\begin{table}[!htbp]
\caption{Time and Token Consuming}
\begin{center}
\begin{tabular}{c c c}
\hline 
\textbf{Method}&\textbf{ATC}&\textbf{MTC}\\
\hline
         \textbf{KnowledgeMind}&134&4362\\
         COT&30&15786\\
         mABC&94&34348\\
         RCAgent&85&29324\\
\hline
\hline
\end{tabular}
\label{tab:acc}
\vspace{-10pt}
\end{center}
\end{table}

\begin{table*}[!htbp]
\caption{Ablation Study of KnowledgeMind}
\begin{center}
\begin{tabular}{c c c c c c c}
\hline 
\textbf{Method}&\textbf{FL@1}&\textbf{FT@1}&\textbf{FT@2}&\textbf{FT@3}\\
\hline
         KnowledgeMind(with Fault Case)&0.892&0.677&0.766&0.826&\\
         KnowledgeMind(w/o Fault Case)&0.724&0.509&0.598&0.659&\\
         KnowledgeMind(w/o Metric Agent)&0.180&0.138&0.168&0.180&\\
         KnowledgeMind(w/o Log Agent)&0.611&0.389&0.467&0.497&\\
\hline
\hline
\end{tabular}
\label{tab:ablation}
\end{center}
\end{table*}

\begin{table*}[]
\caption{Performance comparison of different models.The RL is the location accuracy of the root cause and the RA is the accuracy of the type of root cause.}
\begin{center}
\begin{tabular}{c c c c c c c c c c}
\hline 
\textbf{Dataset}&\textbf{Algorithm}&\textbf{LLM Base}&\textbf{Model Category}&\textbf{FL@1}&\textbf{FT@1}&\textbf{FT@2}&\textbf{FT@3}\\
\hline
&&Qwen-QwQ& &0.701&0.503&0.562&0.617\\
&&LLAMA3.3-70B&Dense &0.707&0.491&0.539&0.598\\
&\textbf{KnowledgeMind}&GPT-4-Turbo& &0.718&0.521&0.581&0.659\\
\cline{3-4}
&&DeepSeek&MoE &0.713&0.503&0.574&0.646\\
&&QwenMax& &0.724&0.509&0.598&0.659\\
\cline{2-8}
&&Qwen-QwQ& &0.317&0.251&0.269&0.305\\
&&LLAMA3.3-70B&Dense &0.335&0.275&0.287&0.299\\
&COT&GPT-4-Turbo& &0.365&0.288&0.299&0.323\\
\cline{3-4}
&&DeepSeek&MoE &0.354&0.264&0.281&0.312\\
&&QwenMax& &0.359&0.275&0.299&0.323\\
\cline{2-8}
$\mathcal{A}$&&Qwen-QwQ& &0.485&0.317&0.354&0.407\\
&&LLAMA3.3-70B&Dense &0.473&0.305&0.335&0.396\\
&mABC&GPT-4-Turbo& &0.485&0.330&0.354&0.419\\
\cline{3-4}
&&DeepSeek&MoE &0.479&0.305&0.341&0.413\\
&&QwenMax& &0.473&0.299&0.347&0.413\\
\cline{2-8}
&&Qwen-QwQ& &0.455&0.294&0.329&0.377\\
&&LLAMA3.3-70B&Dense &0.443&0.299&0.353&0.390\\
&RCAgent&GPT-4-Turbo& &0.443&0.312&0.359&0.396\\
\cline{3-4}
&&DeepSeek&MoE &0.455&0.317&0.353&0.396\\
&&QwenMax& &0.467&0.323&0.353&0.401\\

\hline
\hline
&&Qwen-QwQ& &0.898&0.836&0.868&0.891\\
&&LLAMA3.3-70B&Dense &0.879&0.836&0.836&0.868\\
&\textbf{KnowledgeMind}&GPT-4-Turbo& &0.909&0.848&0.868&0.891\\
\cline{3-4}
&&DeepSeek&MoE &0.898&0.836&0.868&0.891\\
&&QwenMax& &0.903&0.836&0.879&0.903\\
\cline{2-8}
&&Qwen-QwQ& &0.752&0.715&0.727&0.752\\
&&LLAMA3.3-70B&Dense &0.727&0.715&0.727&0.727\\
&COT&GPT-4-Turbo& &0.788&0.715&0.758&0.788\\
\cline{3-4}
&&DeepSeek&MoE &0.776&0.715&0.745&0.768\\
&&QwenMax& &0.788&0.727&0.758&0.782\\
\cline{2-8}
$\mathcal{B}$&&Qwen-QwQ& &0.848&0.812&0.830&0.848\\
&&LLAMA3.3-70B&Dense &0.818&0.776&0.788&0.812\\
&mABC&GPT-4-Turbo& &0.836&0.776&0.812&0.824\\
\cline{3-4}
&&DeepSeek&MoE &0.818&0.776&0.818&0.818\\
&&QwenMax& &0.848&0.776&0.818&0.836\\
\cline{2-8}
&&Qwen-QwQ& &0.812&0.776&0.800&0.812\\
&&LLAMA3.3-70B&Dense &0.800&0.776&0.800&0.800\\
&RCAgent&GPT-4-Turbo& &0.824&0.776&0.818&0.824\\
\cline{3-4}
&&DeepSeek&MoE &0.776&0.727&0.758&0.776\\
&&QwenMax& &0.806&0.752&0.776&0.806\\

\hline
\hline
\end{tabular}
\label{tab:acc}
\end{center}
\end{table*}

\subsubsection{The Time and Token Consuming Experiment}
The service-by-service reasoning LLM system designed with knowledge is more complex compared to approaches like RCAAgent and mABC, which perform root cause localization in a single step based on all relevant information using a judge agent. Our method requires multiple calls to the verifier agent to analyze the anomaly levels of each service layer by layer. This approach reduces the demand for LLM context length, but the drawback is that it may require longer reasoning time. Therefore, we use two metrics, \textbf{Average Time Consuming (ATC)} and \textbf{Maximum Token Consumption (MTC)}, to evaluate the time and token overhead. Maximum Token Consumption (MTC) refers to the maximum number of tokens required as input by the model in a single inference step during the fault diagnosis process. This metric is primarily used to assess the current algorithm's demand for the LLM's maximum input token length.

The experiment is evaluated at A dataset. As the table~\ref{tab:ablation} shown, compared to mABC and RCAgent, our method has a higher time overhead but requires fewer tokens per inference. This is because the other algorithms adopt processes like ReAct and CoT, which significantly increase their token consumption per step. These models repeatedly feed previous reasoning processes back into the input for iterative refinement and must tolerate additional token consumption caused by potential reasoning errors from the LLM.

Note that for models like QwenMax, Qwen-QwQ, which have a maximum context length of 32k, we trim the initial part of the context to fit the input. Therefore, the maximum MTC for this model is 32768. The maximum MTC for mABC and RCAgent in the table comes from GPT-4-Turbo.

As the microservice system to be inferred grows larger, our method does not require a significant increase in context window length due to its service-by-service inference process. In contrast, methods like mABC need to evaluate all abnormal services in the system simultaneously, resulting in a linear increase in context window length as the number of services grows. Therefore, their method is not suitable for large-scale microservice systems


\subsubsection{The Ablation Experiment of KnowledgeMind}
Furthermore, we designed ablation experiments to analyze the role of the knowledge base and each agent in our KnowledgeMind framework at Dataset A with Qwen2.5-max.

As the table~\ref{tab:ablation} shown, when KnowledgeMind operates without the history fault cases in knowledge base agent, i.e., performing unsupervised analysis throughout the process, it stills to accurately localize the root cause of some complex failures just by rules. If the Metric Agent is removed, meaning the loss of metric-related information, most fault localization capabilities are lost. This is because almost all failures are reflected as anomalies in the metrics. For example, CPU-related failures can cause abnormal increases in metrics such as $container\_cpu\_cfs\_throttled\_seconds$. Losing the Log Agent also results in a partial loss of failure localization ability, particularly for issues related to increased network latency, packet loss, packet corruption, and I/O failures. Further analysis reveals that these types of failures are primarily characterized by logs.

\section{Conclusion}
In this paper, we propose a knowledge-driven, MCTS-based LLM multi-agent root cause analysis system. Our experimental results demonstrate that, compared to existing large model-based fault localization frameworks, our method achieves better accuracy. Moreover, the service-by-service reasoning process we designed allows the LLM to complete the reasoning of complex systems under token limitations. In future work, we will integrate ReAct more effectively into various agents to enhance their flexibility and autonomous analysis capabilities.

\balance
\normalem

\appendix
\section{IMPLEMENTATION DETAILS}~\label{app:A}
To better reproduce and showcase the details of our work, we provide the prompts of key Agents, along with templates for tool invocation and the case library.
\subsection{Prompt of Verifier Agent}

\begin{verbatim}
You are an operations engineer. The system 
has encountered a failure, and you need to 
score the severity of anomalies for all 
current services based on your existing 
knowledge and the abnormal information 
from all services. The more Metric and 
Log entries associated with a service's 
anomaly, the higher its severity score. 
If both Metric and Log analyses show 
anomalies for a service, it should receive 
a higher score than if only one type 
(Metric or Log) shows anomalies.

The input contains the abnormal behavior of 
all services. The scores should range from 
1 to 8, and the highest score must be 
at least 2 points higher than the other 
scores.

Output format:
<root_service_start>
Service 1
<root_service_end>:

<root_score_start>
9
<root_score_end>”

Your output must include a step-by-step 
process with detailed analysis and 
reasoning, clearly explaining why 
the top-scoring service received 
the highest score.
\end{verbatim} 

\subsection{Prompt of Metric Agent}
\begin{verbatim}
You are an intelligent metric analysis 
assistant with the ability to perform 
in-depth analysis of Metrics using tools. 
Your task is to identify potential 
anomalies based on user requirements 
and reasonably infer their causes. 

Thinking process:
-Determine the Metric to be analyzed 
(e.g., CPU usage, memory 
consumption, network throughput, etc.).
-Check whether a specific time range or 
target service is involved. 
-Combine historical data and context, 
and use tools to obtain actual analysis 
results.
-Provide the analysis results.

Please follow the format below for 
tool invocation:
<\startoftool>
Tool Name: metric_threshold_analysis_tool  
Input Parameter:  
- metric_name: 
[choose the specific metric]  
- service_name: 
[choose the specific service]  
<\endoftool>

Available tools are as follows:

Tool Name: metric_threshold_analysis_tool  
Input Parameter:  
- metric_name: 
[choose the specific checked metric]  
- service_name: 
[choose the specific checked service]  

Tool Name: metric_timeseries_analysis_tool  
Input Parameter:  
- algorithm_name: 
[choose the specific timeserie algorithm]  
- service_name: 
[choose the specific checked service]  

After obtaining the results, you need 
to diagnose the data based 
on the returned information 
and clearly specify: 

Whether an anomaly was detected 
and the exact values of the abnormal
metrics and the possible causes
\end{verbatim} 

\subsection{Prompt of Summary}
\begin{verbatim}
You are an operations engineer. 
The system has encountered a failure, 
and you need to assess the possible 
types of anomalies based on 
your existing knowledge and the abnormal 
information from the current service. 
The more Metric and Log entries 
associated with a particular type of 
anomaly, the more severe that type of 
failure is likely to be. For example, 
if there are 10 abnormal 
CPU metric entries, the top 1 anomaly 
would be a CPU failure.

Existing knowledge:

Logs are enclosed within 
<\startofloglibrary> 
and 
<\endofloglibrary>.
Metrics are enclosed within 
<\startofmetriclibrary> 
and 
<\endofmetriclibrary>.
Output format:

“top 1:
<root_type_start>
CPU failure, detailed failure analysis...
<root_type_end>”
“top 2:
<root_type_start>
Detailed failure analysis...
<root_type_end>”
“top 3:
<root_type_start>
Detailed failure analysis...
<root_type_end>”

Your output must follow a 
step-by-step process, 
including detailed analysis and 
reasoning. If top 2 and top 3 
cannot be determined, 
only top 1 should be provided.
\end{verbatim}

\subsection{Prompt of MCTS's Simulation}
\begin{verbatim}
You are an operations engineer, and 
now the system has encountered 
a failure. Based on your existing 
knowledge, you need to assess 
the severity of the current 
service's anomaly. The score 
should be determined by both 
the current service's abnormal 
behavior and the impact it has 
on the surrounding services.

The input includes the abnormal behavior 
of the current service and 
all the abnormal behaviors of 
the services calling this service. 
The stronger the abnormal behavior 
of the surrounding services, 
the higher the score should be. 
The score should range from 1 to 10.

The output format must follow this structure:
"1. The score of the current service is:
<root_score_start> 
a number between 1 and 10
<root_score_end>

<root_cause_start>
The detailed analysis process
<root_cause_end>"

Your output must be a complete 
step-by-step process, including 
detailed analysis, reasoning, 
and the final choice of the score.
\end{verbatim} 

\subsection{Prompt of Service-Pod Agent}
\begin{verbatim}
You are an experienced operations engineer, 
and the system is currently experiencing a 
failure. Your task is to use your existing 
knowledge, combined with 
the summary information provided by multiple 
foundational Agents, to analyze and identify 
the root cause service or node, and 
determine the detailed type of failure.

Please note:

1.The root cause may lie in a pod, service, 
or node. For example, adservice is a service; 
adservice-0, adservice-1, adservice-2, and 
adservice2-0 are four pods. Be careful not to 
misidentify the specific root cause.

2.Relationship between services and pods: 
Each service contains multiple pods. 
For instance, adservice may have adservice-0, 
adservice-1, adservice-2, and adservice2-0 
under it.

3.If multiple pods under the same service 
exhibit anomalies, the service is more likely 
to be the root cause rather than an 
individual pod.

4. Summary information from 
multiple foundational Agents to further analyze 
the root cause and type of failure.
\end{verbatim}
\bibliographystyle{IEEEtran}
\bibliography{DSN25}

\begin{thebibliography}{10}
\providecommand{\url}[1]{#1}
\csname url@samestyle\endcsname
\providecommand{\newblock}{\relax}
\providecommand{\bibinfo}[2]{#2}
\providecommand{\BIBentrySTDinterwordspacing}{\spaceskip=0pt\relax}
\providecommand{\BIBentryALTinterwordstretchfactor}{4}
\providecommand{\BIBentryALTinterwordspacing}{\spaceskip=\fontdimen2\font plus
\BIBentryALTinterwordstretchfactor\fontdimen3\font minus \fontdimen4\font\relax}
\providecommand{\BIBforeignlanguage}[2]{{%
\expandafter\ifx\csname l@#1\endcsname\relax
\typeout{** WARNING: IEEEtran.bst: No hyphenation pattern has been}%
\typeout{** loaded for the language `#1'. Using the pattern for}%
\typeout{** the default language instead.}%
\else
\language=\csname l@#1\endcsname
\fi
#2}}
\providecommand{\BIBdecl}{\relax}
\BIBdecl

\bibitem{seer}
Y.~Gan, Y.~Zhang \emph{et~al.}, ``Seer: Leveraging big data to navigate the complexity of performance debugging in cloud microservices,'' in \emph{Proceedings of the twenty-fourth international conference on architectural support for programming languages and operating systems}, 2019, pp. 19--33.

\bibitem{deathstar}
\BIBentryALTinterwordspacing
------, ``An open-source benchmark suite for microservices and their hardware-software implications for cloud {\&} edge systems,'' in \emph{Proceedings of the Twenty-Fourth International Conference on Architectural Support for Programming Languages and Operating Systems, {ASPLOS} 2019, Providence, RI, USA, April 13-17, 2019}, I.~Bahar, M.~Herlihy, E.~Witchel, and A.~R. Lebeck, Eds.\hskip 1em plus 0.5em minus 0.4em\relax {ACM}, 2019, pp. 3--18. [Online]. Available: \url{https://doi.org/10.1145/3297858.3304013}
\BIBentrySTDinterwordspacing

\bibitem{suite}
A.~Sriraman and T.~F. Wenisch, ``$\mu$ suite: a benchmark suite for microservices,'' in \emph{2018 IEEE International Symposium on Workload Characterization (IISWC)}.\hskip 1em plus 0.5em minus 0.4em\relax IEEE, 2018, pp. 1--12.

\bibitem{intro1}
{\'A}.~Brand{\'o}n, M.~Sol{\'e}, A.~Hu{\'e}lamo, D.~Solans, M.~S. P{\'e}rez, and V.~Munt{\'e}s-Mulero, ``Graph-based root cause analysis for service-oriented and microservice architectures,'' \emph{Journal of Systems and Software}, vol. 159, p. 110432, 2020.

\bibitem{ver3}
\BIBentryALTinterwordspacing
B.~Beyer, C.~Jones, J.~Petoff, and N.~R. Murphy, \emph{Site Reliability Engineering: How Google Runs Production Systems}, 2016. [Online]. Available: \url{http://landing.google.com/sre/book.html}
\BIBentrySTDinterwordspacing

\bibitem{icse3}
Y.~Lou, Q.~Zhu, J.~Dong, X.~Li, Z.~Sun, D.~Hao, L.~Zhang, and L.~Zhang, ``Boosting coverage-based fault localization via graph-based representation learning,'' in \emph{Proceedings of the 29th ACM Joint Meeting on European Software Engineering Conference and Symposium on the Foundations of Software Engineering}, 2021, pp. 664--676.

\bibitem{icse4}
M.~Zeng, Y.~Wu, Z.~Ye, Y.~Xiong, X.~Zhang, and L.~Zhang, ``Fault localization via efficient probabilistic modeling of program semantics,'' in \emph{Proceedings of the 44th International Conference on Software Engineering}, 2022, pp. 958--969.

\bibitem{icse9}
L.~Wu, J.~Tordsson, J.~Bogatinovski, E.~Elmroth, and O.~Kao, ``Microdiag: Fine-grained performance diagnosis for microservice systems,'' in \emph{2021 IEEE/ACM International Workshop on Cloud Intelligence (CloudIntelligence)}.\hskip 1em plus 0.5em minus 0.4em\relax IEEE, 2021, pp. 31--36.

\bibitem{intro7}
S.~Mehta and R.~Bhagwan, ``Rex: Preventing bugs and misconfiguration in large services using correlated change analysis,'' in \emph{17th USENIX Symposium on Networked Systems Design and Implementation (NSDI 20)}, 2020, pp. 435--448.

\bibitem{intro8}
A.~Mahimkar, ``Rapid detection of maintenance induced changes in service performance,'' in \emph{Proceedings of the Seventh COnference on Emerging Networking EXperiments and Technologies}, 2011, pp. 1--12.

\bibitem{llm3}
Y.~Han, Q.~Du, Y.~Huang, J.~Wu, F.~Tian, and C.~He, ``The potential of one-shot failure root cause analysis: Collaboration of the large language model and small classifier,'' in \emph{Proceedings of the 39th IEEE/ACM International Conference on Automated Software Engineering}, 2024, pp. 931--943.

\bibitem{llm4}
X.~Zhang, S.~Ghosh, C.~Bansal, R.~Wang, M.~Ma, Y.~Kang, and S.~Rajmohan, ``Automated root causing of cloud incidents using in-context learning with gpt-4,'' in \emph{Companion Proceedings of the 32nd ACM International Conference on the Foundations of Software Engineering}, 2024, pp. 266--277.

\bibitem{llm5}
D.~Roy, X.~Zhang, R.~Bhave, C.~Bansal, P.~Las-Casas, R.~Fonseca, and S.~Rajmohan, ``Exploring llm-based agents for root cause analysis,'' in \emph{Companion Proceedings of the 32nd ACM International Conference on the Foundations of Software Engineering}, 2024, pp. 208--219.

\bibitem{llm6}
P.~Las-Casas, A.~G. Kumbhare, R.~Fonseca, and S.~Agarwal, ``Llexus: an ai agent system for incident management,'' \emph{ACM SIGOPS Operating Systems Review}, vol.~58, no.~1, pp. 23--36, 2024.

\bibitem{llm7}
Z.~Ma, A.~R. Chen, D.~J. Kim, T.-H. Chen, and S.~Wang, ``Llmparser: An exploratory study on using large language models for log parsing,'' in \emph{Proceedings of the IEEE/ACM 46th International Conference on Software Engineering}, 2024, pp. 1--13.

\bibitem{Prometheus}
``Prometheus,'' \url{https://prometheus.io/}, 2023.

\bibitem{elasticsearch}
``Elasticsearch,'' \url{https://www.elastic.co/}, 2023.

\bibitem{jaeger}
``Jaeger: open source, end-to-end distributed tracing.'' \url{https://www.jaegertracing.io/}, 2023.

\bibitem{opentelemetry}
``High-quality, ubiquitous, and portable telemetry to enable effective observability.'' \url{https://opentelemetry.io/}, 2023.

\bibitem{monitorassistant}
Z.~Yu, M.~Ma, C.~Zhang, S.~Qin, Y.~Kang, C.~Bansal, S.~Rajmohan, Y.~Dang, C.~Pei, D.~Pei \emph{et~al.}, ``Monitorassistant: Simplifying cloud service monitoring via large language models,'' in \emph{Companion Proceedings of the 32nd ACM International Conference on the Foundations of Software Engineering}, 2024, pp. 38--49.

\bibitem{rcacopilot}
Y.~Chen, H.~Xie, M.~Ma, Y.~Kang, X.~Gao, L.~Shi, Y.~Cao, X.~Gao, H.~Fan, M.~Wen \emph{et~al.}, ``Automatic root cause analysis via large language models for cloud incidents,'' in \emph{Proceedings of the Nineteenth European Conference on Computer Systems}, 2024, pp. 674--688.

\bibitem{rcagent}
Z.~Wang, Z.~Liu, Y.~Zhang, A.~Zhong, J.~Wang, F.~Yin, L.~Fan, L.~Wu, and Q.~Wen, ``Rcagent: Cloud root cause analysis by autonomous agents with tool-augmented large language models,'' in \emph{Proceedings of the 33rd ACM International Conference on Information and Knowledge Management}, 2024, pp. 4966--4974.

\bibitem{mabc}
W.~Zhang, H.~Guo, J.~Yang, Y.~Zhang, C.~Yan, Z.~Tian, H.~Ji, Z.~Li, T.~Li, T.~Zheng \emph{et~al.}, ``mabc: multi-agent blockchain-inspired collaboration for root cause analysis in micro-services architecture,'' \emph{arXiv preprint arXiv:2404.12135}, 2024.

\bibitem{react}
S.~Yao, J.~Zhao, D.~Yu, N.~Du, I.~Shafran, K.~Narasimhan, and Y.~Cao, ``React: Synergizing reasoning and acting in language models,'' \emph{arXiv preprint arXiv:2210.03629}, 2022.

\bibitem{llm1}
P.~Jin, S.~Zhang, M.~Ma, H.~Li, Y.~Kang, L.~Li, Y.~Liu, B.~Qiao, C.~Zhang, P.~Zhao \emph{et~al.}, ``Assess and summarize: Improve outage understanding with large language models,'' in \emph{Proceedings of the 31st ACM Joint European Software Engineering Conference and Symposium on the Foundations of Software Engineering}, 2023, pp. 1657--1668.

\bibitem{llm2}
Y.~Jiang, C.~Zhang, S.~He, Z.~Yang, M.~Ma, S.~Qin, Y.~Kang, Y.~Dang, S.~Rajmohan, Q.~Lin \emph{et~al.}, ``Xpert: Empowering incident management with query recommendations via large language models,'' in \emph{Proceedings of the IEEE/ACM 46th International Conference on Software Engineering}, 2024, pp. 1--13.

\bibitem{drain}
P.~He, J.~Zhu, and Z.~Zheng, ``Drain: An online log parsing approach with fixed depth tree,'' in \emph{2017 IEEE international conference on web services (ICWS)}.\hskip 1em plus 0.5em minus 0.4em\relax IEEE, 2017, pp. 33--40.

\bibitem{AIOPS2022}
``Aiops 2022 championship,'' \url{https://competition.aiops.cn/}, 2022.

\bibitem{gmm}
D.~A. Reynolds \emph{et~al.}, ``Gaussian mixture models.'' \emph{Encyclopedia of biometrics}, vol. 741, no. 659-663, 2009.

\bibitem{tfidf}
K.~Church and W.~Gale, ``Inverse document frequency (idf): A measure of deviations from poisson,'' in \emph{Natural language processing using very large corpora}.\hskip 1em plus 0.5em minus 0.4em\relax Springer, 1999, pp. 283--295.

\bibitem{trainticket}
X.~Zhou, X.~Peng \emph{et~al.}, ``Fault analysis and debugging of microservice systems: Industrial survey, benchmark system, and empirical study,'' \emph{IEEE Transactions on Software Engineering}, vol.~47, no.~2, pp. 243--260, 2018.

\bibitem{dejavu}
Z.~Li, N.~Zhao, M.~Li, X.~Lu, L.~Wang, D.~Chang, X.~Nie, L.~Cao, W.~Zhang, K.~Sui, Y.~Wang, X.~Du, G.~Duan, and D.~Pei, ``Actionable and interpretable fault localization for recurring failures in online service systems,'' in \emph{Proceedings of the 2022 30th {{ACM Joint Meeting}} on {{European Software Engineering Conference}} and {{Symposium}} on the {{Foundations}} of {{Software Engineering}}}, ser. {{ESEC}}/{{FSE}} 2022, Nov. 2022.

\bibitem{sage}
Y.~Gan, M.~Liang \emph{et~al.}, ``Sage: Using unsupervised learning for scalable performance debugging in microservices,'' \emph{arXiv preprint arXiv:2101.00267}, 2021.

\bibitem{slim}
R.~Ren, J.~Yang, L.~Yang, X.~Gu, and L.~Sun, ``Slim: a scalable light-weight root cause analysis for imbalanced data in microservice,'' in \emph{Proceedings of the 2024 IEEE/ACM 46th International Conference on Software Engineering: Companion Proceedings}, 2024, pp. 328--330.

\bibitem{tracerca}
Z.~Li, J.~Chen \emph{et~al.}, ``Practical root cause localization for microservice systems via trace analysis,'' in \emph{2021 IEEE/ACM 29th International Symposium on Quality of Service (IWQOS)}.\hskip 1em plus 0.5em minus 0.4em\relax IEEE, 2021, pp. 1--10.

\bibitem{microrank}
G.~Yu, P.~Chen \emph{et~al.}, ``Microrank: End-to-end latency issue localization with extended spectrum analysis in microservice environments,'' in \emph{Proceedings of the Web Conference 2021}, 2021, pp. 3087--3098.

\bibitem{cot}
J.~Wei, X.~Wang, D.~Schuurmans, M.~Bosma, F.~Xia, E.~Chi, Q.~V. Le, D.~Zhou \emph{et~al.}, ``Chain-of-thought prompting elicits reasoning in large language models,'' \emph{Advances in neural information processing systems}, vol.~35, pp. 24\,824--24\,837, 2022.

\bibitem{moe}
R.~A. Jacobs, M.~I. Jordan, S.~J. Nowlan, and G.~E. Hinton, ``Adaptive mixtures of local experts,'' \emph{Neural computation}, vol.~3, no.~1, pp. 79--87, 1991.

\bibitem{qwen2}
A.~Yang, B.~Yang, B.~Zhang, B.~Hui, B.~Zheng, B.~Yu, C.~Li, D.~Liu, F.~Huang, H.~Wei \emph{et~al.}, ``Qwen2. 5 technical report,'' \emph{arXiv preprint arXiv:2412.15115}, 2024.

\bibitem{llama}
A.~Dubey, A.~Jauhri, A.~Pandey, A.~Kadian, A.~Al-Dahle, A.~Letman, A.~Mathur, A.~Schelten, A.~Yang, A.~Fan \emph{et~al.}, ``The llama 3 herd of models,'' \emph{arXiv preprint arXiv:2407.21783}, 2024.

\bibitem{gpt4}
J.~Achiam, S.~Adler, S.~Agarwal, L.~Ahmad, I.~Akkaya, F.~L. Aleman, D.~Almeida, J.~Altenschmidt, S.~Altman, S.~Anadkat \emph{et~al.}, ``Gpt-4 technical report,'' \emph{arXiv preprint arXiv:2303.08774}, 2023.

\bibitem{qwenmax}
``Qwen max,'' \url{https://qwenlm.github.io/blog/qwen2.5-max/}, 2022.

\bibitem{deepseek}
A.~Liu, B.~Feng, B.~Xue, B.~Wang, B.~Wu, C.~Lu, C.~Zhao, C.~Deng, C.~Zhang, C.~Ruan \emph{et~al.}, ``Deepseek-v3 technical report,'' \emph{arXiv preprint arXiv:2412.19437}, 2024.

\end{thebibliography}

\end{document}